\documentclass[12pt,a4paper]{article}
\usepackage[latin1] {inputenc}
\usepackage{amsmath}
\usepackage{amsfonts}
\usepackage{amssymb}
\usepackage[dvips]{graphicx,color}
\title{On Coordinate Transformations in Planar Noncommutative Theories}
\author{
D.~H.~Correa$^a$\thanks{CONICET} \,, C.~D.~Fosco$^{b *}$ \,,\\
F.~A.~Schaposnik$^a$\thanks{Associated with CICPBA}\;,
and G.~Torroba$^{b}$  \\
{\normalsize\it  $^a$Departamento de F\'\i sica, Universidad
Nacional
de La Plata}\\
{\normalsize\it C.C. 67, 1900 La Plata, Argentina}
\\
{\normalsize\it $^b$Centro At\'omico de Bariloche and Instituto
Balseiro}\\
{\normalsize\it Comisi\'on Nacional de Energ\'\i a At\'omica}
\\
{\normalsize\it 8400 Bariloche, Argentina} }
\newcommand{\hp}{\,\hat{\!\begin{eqnarray}r \phi}}

\newcommand{\es}{{\begin{eqnarray}r*}}
\begin{document}
%
\date{}
\maketitle
\begin{abstract}
\noindent We consider planar noncommutative theories such that the
coordinates verify a space-dependent commutation relation. We show
that, in some special cases, new coordinates may be introduced
that have a constant commutator, and as a consequence the
construction of Field Theory models may be carried out by an
application of the standard Moyal approach in terms of the new
coordinates.

We apply these ideas to the concrete example of a noncommutative
plane with a curved interface. We also show how to extend this
method to more general situations.
\end{abstract}


\section{Introduction}\label{sec:intro}
Noncommutativity has been intensively studied in past years, in
part due to its role in String Theory (\cite{string} and
references therein) and its application to Condensed Matter
Physics (see, for example, \cite{fosco} and references therein).
It also offers interesting alternatives for gauge theories. From
the mathematical point of view, the development of Noncommutative
Geometry~\cite{connes} and the Formality Theorem in Deformation
Quantization~\cite{ko1}, have been fundamental to the
understanding of noncommutativity.

The case of a constant noncommutativity parameter, which leads to
the familiar Moyal product, is well understood (see, for example,
\cite{DN}). However, space-dependent noncommutativity is much more
involved, since there is no general procedure to find an explicit
representation for the resulting infinite-dimensional algebra,
except when it comes from a Poisson structure~\cite{ko1}. The
whole physical picture corresponding to this situation is,
consequently, not completely known, although a lot of progress has
been made due to recent efforts. For example, the connection
between curved D-branes and space-dependent noncommutativity has
been established in ~\cite{cornalba}, and the construction of
gauge theories in curved noncommutative spaces has been presented
in~\cite{gauge}. See also refs.\cite{Jack1}-\cite{Amelino}.
Besides, an illuminating field-theory interpretation of
Kontsevich\'{}s construction in terms of a sigma model has been
provided in~\cite{cf1}.

The aim of this paper is to show how certain cases of coordinates
satisfying a space-dependent commutation relation may be analyzed
with the help of a change of variables. The main idea is that it
may be possible to define new variables in terms of which one
obtains a constant commutator. In terms of those coordinates, we
can work with the usual Moyal representation of the noncommutative
product and  then easily construct a calculus and study the
geometry of the space. If the relevant objects (product,
derivatives, integrals,\ldots) may at the end be written in terms
of the original variables, we  can obtain explicit representations
for the original algebra, as well as for the derivatives and
integrals. Interestingly enough, cases in which this can be
achieved are not artificial but arise in a variety of problems,
for example related to the construction of noncommutative solitons
and instantons \cite{sc}-\cite{sc1}.

This paper is organized as follows: in section~\ref{sec:metric},
we discuss the relation between the metric and a space-dependent
commutation relation for the coordinates of a two-dimensional
space, based on general considerations. In
section~\ref{ssec:particular}, we consider the particular case of
a commutation relation that depends on only one of the
coordinates. We show that, for this particular case, it is always
possible to perform a change of variables to new coordinates
verifying a constant (i.e., space-independent) commutation
relation so that the Moyal product representation can be used with
the resulting simplifications it implies. We then extend all the
previous results to the example of a curved interface dividing two
regions that have (different) constant values for $\theta$ in
section~\ref{sec:interface}, where we present the concrete example
of a scalar field theory. The possibility of generalizing the
method is discussed in section~\ref{sec:general}. In
section~\ref{sec:conc} we present our conclusions.


\section{Noncommutative products and two dimensional metrics}\label{sec:metric}
Coordinate commutators in a non-flat background spatial metric,
necessarily imply the introduction of a coordinate-dependent
noncommutativity parameter $\theta^{ij}$:
 \begin{equation}\label{uno}
 [x^i, x^j]_\star = i\theta^{ij}(x) \;.
\end{equation}
Now, as it is well known, there are severe constraints that have
to be satisfied for a consistent definition of a noncommutative
associative $\star$-product. In general, even when those
constraints are satisfied, it is not possible to give a closed and
explicit formula for the $\star$-product between two arbitrary
functions, as it can be done for the constant-$\theta$ case (where
one has the standard Moyal product). The general conditions under
which such a product can be defined are derived in~\cite{ko1}.

A sufficient condition to have an associative noncommutative
 product may be stated as follows ~\cite{sc}:
\begin{equation}\label{suff}
\nabla_i \theta^{jk}(x) = 0 \;,
\end{equation}
where $\nabla_i$ is the  covariant derivative corresponding to a
metric connection. Although up to this point we have not introduced any
metric into the game, a  natural one  will appear precisely when solving
(\ref{suff}) in $2$ dimensions. That metric  will be determined (up to a constant scalar factor)  by $\theta^{jk}$ itself; see (\ref{conchap1})
and (\ref{conchap2}) below.

Condition (\ref{suff}) may be derived by starting from the
definition of the Poisson bracket $\{f,h\}$ for two functions $f$
and $h$:
\begin{equation}\label{pb}
\{f,h\} \; = \; \partial_i f\,  \theta^{ij} \, \partial_j h
\end{equation}
and then noting that a noncommutative product, up to order
$\theta$, takes the form~\footnote{Higher orders in an expansion
in powers of $\theta$ are presented
  in~\cite{ko1}.}:
\begin{equation}\label{starr}
f \star h \equiv fh + \frac{i}{2}\{f,h\} + {\cal O}(\theta^2)\;.
\end{equation}

The associativity of the $\star$-product (\ref{starr}) requires
the Jacobi identity for the Poisson bracket (\ref{pb}) to hold
true, which in turns results in the equation:
\begin{equation}
\theta^{ij}\partial_j\theta^{kl} +
\theta^{lj}\partial_j\theta^{ik} +
\theta^{kj}\partial_j\theta^{li} = 0 \;,
\end{equation}
i.e., $\theta^{ij}$ is a Poisson structure. In fact, this
implies~\cite{cornalba} that the product will be associative to
all orders.

The previous condition can be written covariantly in the form:
\begin{equation}
\label{eq:jacovar} \theta^{ij}\nabla_j\theta^{kl} +
\theta^{lj}\nabla_j\theta^{ik} + \theta^{kj}\nabla_j\theta^{li} =
0 \;,
\end{equation}
since the terms in $\nabla_j$ containing the (symmetric) connection
cancel out. Condition (\ref{suff}) is a
sufficient condition for (\ref{eq:jacovar}) to be true, namely, for
the associativity to be valid to all orders in $\theta$.

Moreover, (\ref{suff}) is very easy to deal with on a two
dimensional space. Indeed, the most general $\theta^{ij}$ can be
written in $d=2$ in the form
\begin{equation}
 \theta^{ij} (x^1,x^2) = \frac{\varepsilon^{ij}}{\sqrt {g(x)}} \theta_0(x^1,x^2)
\end{equation}
where $\theta_0(x^1,x^2)$ is a scalar and $g(x)$ is the
determinant of the two dimensional metric $g_{jk}(x)$. But then,
condition (\ref{suff}) reduces to
\begin{equation}
\nabla_i  \left(\frac{\varepsilon^{jk}}{\sqrt{g(x)}} \theta_0(x^1,x^2) \right)= \frac{\varepsilon^{jk}}{\sqrt{g(x)}}
\nabla_i \theta_0(x^1,x^2) = \frac{\varepsilon^{jk}}{\sqrt {g(x)} }\partial_i\theta_0(x^1,x^2) = 0
\label{conchap}
\end{equation}
and hence $\theta_0$ must be a constant.

Then, the sufficient condition (\ref{suff}) is equivalent, in $d=2$,
to the equation:
\begin{equation}
\theta^{ij}(x^1,x^2) = \frac{\varepsilon^{ij}}{\sqrt {g(x)}} \, \theta_0 \;,
\end{equation}
where $\theta_0$ is a real (non-vanishing) constant. This
expression relates the  space-dependent associative
noncommutativity to a non-trivial background metric $g_{ij}$. In
fact, it is believed~\cite{madore} that quantum gravity is at the
origin of noncommutative effects, so that kind of relation is not
entirely unexpected. The $g_{ij}$ tensor can be explicitly written
by taking into account that in $2$ dimensions every metric is
conformally flat:
\begin{equation}\label{conchap1}
g_{ij} (x) \;=\; e^{\sigma (x)} \; \delta_{ij} \;,
\end{equation}
and that the conformal factor $e^{\sigma (x)}$ is determined by $g$:
\begin{equation}\label{conchap2}
e^{\sigma (x)} \;=\; \sqrt{ g(x)} \;.
\end{equation}

Let us also note that, in the presence of a metric, the
natural integration measure, $d\mu$,  should be
\begin{equation}\label{dmu}
 d\mu \;=\;  d^2 x \sqrt {g(x)}
\end{equation}
which is, as we shall see, consistent with the
definition of a noncommutative product in which the integral acts as a
trace.

\subsection{The case $\theta = \theta(x^1)$}\label{ssec:particular}
Let us specialize the previous discussion in two dimensional space
to the case in which $\theta$ depends on only one coordinate,
$x^1$ say. We know that $\theta$ can then be written as follows:
\begin{equation}
\theta \;=\;  \theta_0 \, t(x^1) \;,
\end{equation}
where $t(x^1)\;=\; 1/ \sqrt{g(x)}$ is a positive definite
function.

Our aim is to present an explicit formula for  a noncommutative
associative  $\star$-product such that
 \begin{equation}
[x^1, x^2]_{ \star} \;=\;  x^1 \star  x^2  - x^2 \star  x^1  \;=\;
i \, \theta_0 \, t(x^1) \label{11} \;.
\end{equation}
An expression for $\star$ in powers of $\theta_0$ was already
found by Kontsevich \cite{ko1}, where the first terms are given by
(\ref{starr}). However, the procedure to construct it, based on
the Formality Theorem, can be quite involved in practice. Rather
than following that approach, we shall show how to arrive to
 an associative noncommutative product by a different path. Inspired by the change of
variables found in~\cite{sc}, while studying noncommutative
vortices in a curved space, we left and right multiply (\ref{11})
by $1/{\sqrt{t(x^1)}}$ where the square root is the noncommutative
one, i.e., $\sqrt{a} \star \sqrt{a} = a$. Then:
\begin{equation}
x^1 \star \frac{1}{\sqrt{t(x^1)}} \star x^2 \star
\frac{1}{\sqrt{t(x^1)}}-
 \frac{1}{\sqrt{t(x^1)}} \star x^2 \star  \frac{1}{\sqrt{t(x^1)}}
\star x^1 = i \theta_0 \;.
\label{2}
\end{equation}
Here we have used the associativity of the $\star$-product, which
is valid for any positive function $t(x^1)$.

 We then change variables, from $(x^1,x^2)$ to new
ones $(y^1, y^2)$ defined by:
\begin{equation}\label{tres}
\left\{
\begin{array}{ccl}
y^1 &=& x^1 \\
y^2 &=&  \frac{1}{\sqrt{t(x^1)}} \star x^2 \star
\frac{1}{\sqrt{t(x^1)}}
\end{array}
\right.
\end{equation}
obtaining:
\begin{equation}
y^1  \star  y^2 \,-\,  y^2 \star y^1 \;=\; i \, \theta_0 \;.
\label{cuatro}
\end{equation}
Note that (\ref{tres}) involves the (well-defined) square root of
a positive element, and besides, the definition of the new
variables is not sensitive to the sign ambiguity of the square
root, since that square root appears quadratically. Also, the
variables $y^1$ and $y^2$, as defined in (\ref{tres}), are
Hermitian and, because of (\ref{cuatro}), the noncommutative
$\star$-product can then be realized as an ordinary Moyal product.
We shall use `$*$' to denote the constant $\theta$ ($=\theta_0$)
Moyal product in terms of variables $(y^1,y^2)$:
\begin{equation}\label{eq:moyal1}
f(y) \star g(y)\,=\, f(y)*
g(y)\,=\,{\rm{exp}}\big(\frac{i}{2}\theta_0
\varepsilon^{jk}\,\frac{\partial}{\partial y
^j}\frac{\partial}{\partial \tilde y ^k} \big)\,f(y) g(\tilde y)
\; \Big \vert _{\tilde y = y} \;.
\end{equation}
For functions of $y^1$ and $y^2$, working  with the $*$-product,
we can use all the standard (flat-space) noncommutative geometry
tools \cite{DN}.

Besides, we can rewrite the Moyal formula in terms of the original
variables,  where it leads to a concrete expression for the
$\star$-product. To that end, we first note that from (\ref{tres})
one has
\begin{equation}
\left\{
\begin{array}{ccl}
x^1 &=& y^1 \\
x^2 &=& \sqrt{t(y^1)} \ast y^2 * \sqrt{t(y^1)}\;.
\end{array}
\right.
\end{equation}
Using (\ref{eq:moyal1}), we obtain

\begin{equation}
\sqrt{t(y^1)} \ast y^2 * \sqrt{t(y^1)} \;=\; t(y^1) \; y^2 \;\;\;.
\end{equation}
where all the noncommutative artifacts have disappeared. Then:
\begin{equation}\label{eq:changecoord}
\left\{
\begin{array}{ccl}
x^1 &=& y^1 \\
x^2 &=& t(y^1)\; y^2 \;,
\end{array}
\right.
\end{equation}
is an (exact) expression for the change of variables, that will allow
us to derive explicit expressions for $\star$ and for the integration measure
in the `physical' variables $x^1$, $x^2$.

As an example, consider the $\star$-product between two functions
of the original variables $x^1,x^2$. It can be defined in the form
\begin{eqnarray}
f(x^1,x^2) \star g(x^1,x^2)\!\!\! &\equiv& \!\!\! f(y^1,t(y^1)y^2)
* g(y^1,t(y^1)y^2)
\nonumber\\
 \!\!\! &=& \!\!\!
{\rm{exp}}\big(\frac{i}{2}\theta_0
\varepsilon^{jk}\,\frac{\partial}{\partial y
^j}\frac{\partial}{\partial \tilde y ^k} \big)\,f(y^1, t(y^1) y^2)
\, g(\tilde y ^1, t(\tilde y ^1) \tilde y ^2) \; \Big \vert
_{\tilde y = y}.\nonumber\\
 \label{cinco}
\end{eqnarray}
It is easy to verify that (\ref{cinco}) provides a consistent
associative realization of the algebra (\ref{11}).

The first line in  eq.(\ref{cinco}) may be applied to verify {\em
  a posteriori\/} the associativity of the $\star$-product to all orders.
Indeed, that property is, in this light, simply {\em inherited\/} from
the (well-known) associativity of the $\ast$ product:
$$
\big( \,f(x) \star g(x) \, \big) \star h(x) \;=\; \big( \, f(x(y))
* g(x(y)) \,\big) \ast h(x(y))
$$
\begin{equation}
=\; f(x(y)) * \big(\, g(x(y)) \ast h(x(y)) \,\big) \;=\; f(x(y))
\star \big(\, g(x(y)) \star h(x(y)) \,\big) \;.
\end{equation}

Of course, since we have at our disposal the formula for the
change of variables (\ref{eq:changecoord}), we may write
everything in terms of $x^1$ and $x^2$ on the right hand side of
eq.(22). One has to take into account the formul{\ae}
\begin{eqnarray}
\left. \frac{\partial}{\partial y^1}\right\vert_{y^2} &=& \left.
\frac{\partial}{\partial x^1}\right\vert_{x^2} +
x^2\frac{t'(x^1)}{t(x^1)} \left. \frac{\partial}{\partial
x^2}\right\vert_{x^1} \nonumber\\
\left. \frac{\partial}{\partial y^2}\right\vert_{y^1} &=& t(x^1)
\left. \frac{\partial}{\partial x^2} \right \vert_{x^1}
\label{car}
\end{eqnarray}
Using (\ref{car}), the $\star$-product of functions of the
original variables can be finally written, to all orders in
$\theta_0$, as
\begin{eqnarray}
\label{eq:moyal2} f(x)\! \star \!g(x)&\!\! = \!\!&
{\rm{exp}}\frac{i \theta_0 }{2}\left[t(\tilde
x^1)\frac{\partial}{\partial x^1}\frac{\partial}{\partial \tilde
x^2}-t(x ^1)\frac{\partial}{\partial x^2}\frac{\partial}{\partial
\tilde x^1}\right. \nonumber\\
&& \,\,\,\,\,\,\,\,\,\,\,\,\,\,\,\,\,\,\,\left.
+\!\left(\!x^2\frac{t'(x^1\!)}{t(x^1\!)}t(\tilde x^1\!)- \tilde
x^2\frac{t'(\tilde x^1\!)}{t(\tilde x^1\!)}t(x^1\!)\!
\right)\!\frac{\partial}{\partial
x^2}\frac{\partial}{\partial\tilde x^2}
 \right]\!f(x) g(\tilde x) \Big \vert _{\tilde x = x}\nonumber\\
\end{eqnarray}
By explicit computation order to order in $\theta_0$ one can
verify the {\it inherited} associativity.
For example, up to order
$\theta_0^2$, the product (\ref{eq:moyal2}) takes the form
\begin{eqnarray}
\label{eq:moyal3} f(x) \star g(x)&\! = \!& f(x)g(x)
+\frac{i\theta_0 t(x^1)}{2}\left(\partial_1 f
{\partial_2 g}- {\partial_2 f}
{\partial_1 g}\right)
\nonumber\\
&&-\frac{\theta_0^2 t(x^1)^2}{8}\left(\partial_1^2f\partial_2^2g +
 \partial_2^2f\partial_1^2g
-2 \partial_1\partial_2f\partial_1\partial_2g\right)
\nonumber\\
&& -\frac{\theta_0^2}{8}\left( x^2 t(x^1)t(x^1)''
\partial_2\left(\partial_2f\partial_2g\right) -2 t'(x^1)^2
\partial_2\left(x^2 \partial_2f\partial_2g\right)
\right)
\nonumber\\
&& +\frac{\theta_0^2}{4}\left(
t(x^1)t'(x^1)\partial_1\left(\partial_2f\partial_2g\right)\right)
+{\cal O}(\theta_0^3)
\end{eqnarray}
where $\partial_1$ and $\partial_2$ are partial derivatives with respect
to $x^1$ and $x^2$. One can explicitly verify that this formula satisfies
associativity.

Note that Kontsevich construction
 \cite{ko1}
leads to equivalence classes (one for each particular function $\theta$)
of associative products. The product we have
defined is not the representant chosen in \cite{ko1} where the procedure
to change the representant is given.

We are   ready now to consider integrals of products of fields
and derivatives, starting from the known expressions in terms of
the new variables. Taking into account that:
\begin{equation}
dy^1 \, dy^2 \;=\; dx^1 \, dx^2 \; t^{-1}(x^1) \;.
\end{equation}
and starting from an integral $I_2$ involving the $\ast$-product
of two functions, we can write  an equivalent expression in the
new variables:
\begin{equation}
I_2 = \int dy^1 dy^2 \tilde{f}(y^1,y^2) * \tilde{g}(y^1,y^2) \;=\;
\int dx^1 dx^2 \frac{1}{t(x^1)} f(x^1,x^2)\star g(x^1,x^2) \;,
\end{equation}
where $\tilde f (y) \equiv f \big(x(y) \big)$.  This expression
guarantees that
 integrals of quadratic terms  coincide with ordinary ones
(when fields satisfy appropriate boundary conditions). Up to order
$\theta^2$ this can be verified using eq.(\ref{eq:moyal3}).

 Hence we see that also in this approach, the natural
integration measure, in terms of variables $(x^1,x^2)$ is that
given in (\ref{dmu}),
\begin{equation}
d\mu(x) = dx^1 dx^2  t^{-1}(x^1) \;.
\end{equation}
This agrees with the result obtained in~\cite{ft}, based on an
operatorial description of noncommutativity.

Let us end the previous discussion by noting that the machinery
developed in this section is very helpful in the construction of
solitons (vortices and monopoles) and instantons in noncommutative
gauge theories.
 Indeed, one can connect an axially symmetric (in time
 $t$) ansatz for instantons in $4$-dimensional noncommutative
 Yang-Mills theory with noncommutative vortices in two dimensional
 curved space with variables $(r,t)$ . Now, the metric in which this
 connection is realized takes the form $g^{ij} = r^2 \delta^{ij}$. Since it
 depends on just one variable, one can proceed, as explained above, to
 change variables in order to work with a standard Moyal product. In
 this way, passing as usual to the Fock space framework, explicit
 vortex, monopole and instanton solutions can be found in a very
 simple way~\cite{sc}-\cite{sc1}.

\section{Example: interface between regions with different
$\theta$.}\label{sec:interface}

As an application (and extension) of the previous results, we
shall study now the case in which $\theta(x^1,x^2)$ is a function
whose variation is concentrated along a curve ${\mathcal C}$, and
is approximately constant elsewhere. In particular, we are
interested in considering cases where $\theta$ has different
(constant) values on two different regions which have ${\mathcal
C}$ as a common interface.

\begin{figure}[h]
\begin{center}
\begin{picture}(0,0)%
\includegraphics{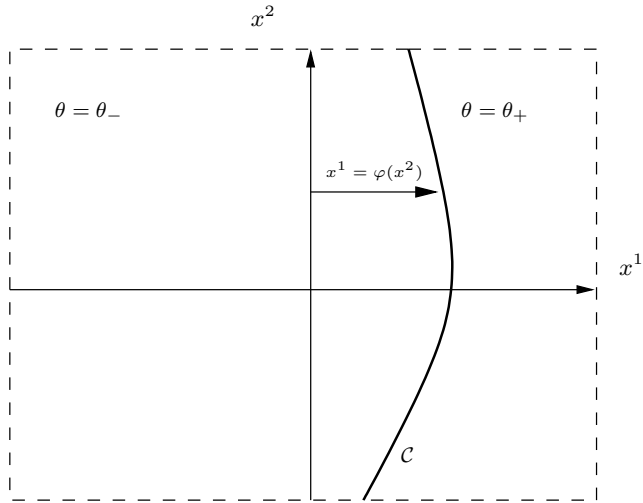}%
\end{picture}%
\setlength{\unitlength}{4144sp}%
\begingroup\makeatletter\ifx\SetFigFont\undefined%
\gdef\SetFigFont#1#2#3#4#5{%
  \reset@font\fontsize{#1}{#2pt}%
  \fontfamily{#3}\fontseries{#4}\fontshape{#5}%
  \selectfont}%
\fi\endgroup%
\begin{picture}(3657,2953)(889,-2513)
\put(2791,-556){\makebox(0,0)[lb]{\smash{\SetFigFont{6}{7.2}{\rmdefault}{\mddefault}{\updefault}{\color[rgb]{0,0,0}$x^1 = \varphi(x^2)$}%
}}}
\put(4546,-1141){\makebox(0,0)[lb]{\smash{\SetFigFont{9}{10.8}{\familydefault}{\mddefault}{\updefault}{\color[rgb]{0,0,0}$x^1$}%
}}}
\put(2341,344){\makebox(0,0)[lb]{\smash{\SetFigFont{9}{10.8}{\familydefault}{\mddefault}{\updefault}{\color[rgb]{0,0,0}$x^2$}%
}}}
\put(1171,-196){\makebox(0,0)[lb]{\smash{\SetFigFont{8}{9.6}{\familydefault}{\mddefault}{\updefault}{\color[rgb]{0,0,0}$\theta = \theta_-$}%
}}}
\put(3241,-2266){\makebox(0,0)[lb]{\smash{\SetFigFont{8}{9.6}{\familydefault}{\mddefault}{\updefault}{\color[rgb]{0,0,0}${\mathcal C}$}%
}}}
\put(3601,-196){\makebox(0,0)[lb]{\smash{\SetFigFont{8}{9.6}{\familydefault}{\mddefault}{\updefault}{\color[rgb]{0,0,0}$\theta = \theta_+$}%
}}}
\end{picture}
\caption{Example of a $\theta$ that takes the values $\theta_-$ and $\theta_+$ in two regions
which are separated by the curve ${\mathcal C}$}
\label{fig:single}
\end{center}
\end{figure}

A simple, and yet non-trivial example of this situation corresponds to
a  $\theta (x^1,x^2)$ with the following structure:
\begin{equation}\label{eq:single}
\theta (x^1,x^2) \;=\; \theta_+ \, H(x^1 - \varphi(x^2)) \,+\,
\theta_- \, H(\varphi(x^2) - x^1)
\end{equation}
where $\theta_+$ and $\theta_-$ are two different real constants,
and $H$ denotes the Heaviside's step function, with integral
representation
\begin{equation}\label{eq:defh}
H(x) \;=\; \int_{-\infty}^{+\infty} \frac{d \nu}{2\pi i} \,
\frac{e^{i \nu  x}}{\nu - i 0^+}\;.
\end{equation}
The shape of the interface is determined by the function $\varphi$, which we
assume to be smooth. The zeroes in the argument of $H$ correspond to
\begin{equation}\label{eq:ccurve}
x^1 - \varphi (x^2) \;=\; 0 \;,
\end{equation}
which defines a smooth curve ${\mathcal C}$, dividing the plane into
two regions with different values of $\theta$, denoted $\theta_+$ and $\theta_-$; see
Figure~\ref{fig:single}.

Our first step in constructing a noncommutative theory based on
(\ref{eq:single}) consists in smoothing out the shape of the function
$\theta(x^1,x^2)$, to avoid introducing singularities due to the existence
of a finite jump on the interface~\footnote{The finite jump may be
recovered at the end (if necessary) by considering the appropriate
limit.}.  That effect may be achieved by replacing $H(x)$ by a
smooth approximation to it, $h_\epsilon(x)$, such that $h_\epsilon(x) \to H(x)$ for $\epsilon \to 0$.
For example,
\begin{equation}\label{eq:smoothh}
h_\epsilon(x) \;=\; \frac{1}{2} \, [ 1 + \tanh(\frac{x}{\epsilon}) ] \;.
\end{equation}
With this particular approximant, which we shall adopt,
$\theta(x^1,x^2)$  may be written more explicitly as follows:
\begin{equation}\label{eq:smooththeta}
  \theta(x^1,x^2) \;=\; \theta_0 \,+\, \delta \, \tanh(\frac{x^1 - \varphi(x^2)}{\epsilon})\;,
\end{equation}
where we defined $\theta_0 = \frac{\theta_++\theta_-}{2}$ and $\delta=\frac{\theta_+ - \theta_-}{2}$.

We now proceed to change variables from the original coordinates
$(x^1,x^2)$ to new ones $(Q^1,Q^2)$
which are defined by:
\begin{equation}\label{eq:varchange1}
\left\{
\begin{array}{ccl}
Q^1 &=& x^1 \,-\, \varphi(x^2) \\
Q^2 &=& x^2 \;\;\;,
\end{array}
\right.
\end{equation}
so that the new variables verify the commutation relation:
\begin{equation}\label{eq:commy}
[ Q^1 \,,\, Q^2 ]_\star \;=\; i \,\left(   \theta_0 \,+\, \delta
\, \tanh(\frac{Q^1}{\epsilon}) \right)\;.
\end{equation}
In what follows we shall assume that the constants $\theta_+$ and
$\theta_-$ have the same sign (positive, say), since the opposite
situation, which necessarily involves a passage of $\theta$ through
zero, is qualitatively different, as it will become clear below.  With
this in mind, we may write (\ref{eq:commy}) in the form:
\begin{equation}\label{eq:commy1}
[Q^1 \,,\, Q^2 ]_\star \;=\; i \,\theta_0 \; t(Q^1) \;,
\end{equation}
where:
\begin{equation}
t(Q^1) \;\equiv\; 1 \, +
\,\frac{\delta}{\theta_0}\,\tanh(\frac{Q^1}{\epsilon})\;.
\end{equation}

Assuming that both $\theta_+$ and $\theta_-$ are positive, the
function $t$ shall always be positive, and we are of course in the
same situation we considered in~\ref{ssec:particular}.  Based on
those results, we introduce a second set of variables, $(y^1,
y^2)$, such that:
\begin{equation}\label{eq:varchange2}
\left\{
\begin{array}{ccl}
y^1 &=& Q^1 \\
y^2 &=& \big[ t(Q^1) \big]^{-1/2} \,\star\,  Q^2 \,\star\, \big[
t(Q^1) \big]^{-1/2} \;,
\end{array}
\right.
\end{equation}
which verify the constant-$\theta$ algebra
\begin{equation}\label{eq:commq}
[ y^1 \,,\, y^2 ]_\star \;=\; i \; \theta_0 \;,
\end{equation}
with \mbox{$\theta_0=\frac{\theta_++\theta_-}{2}$} as the noncommutativity
parameter.

Then, in terms of the original variables, we have:
\begin{equation}
\left\{
\begin{array}{ccl}
y^1 &=&  x^1 \,-\, \varphi (x^2) \\
y^2 &=&  \big[ t(x^1 - \varphi(x^2)) \big]^{-1/2} \,\star\, x^2
\,\star\, \big[t(x^1-\varphi(x^2)) \big]^{-1/2}\;\;\;.
\end{array}
\right.
\end{equation}

Of course, by a similar procedure to the one explained
in~\ref{ssec:particular}, we may write everything in terms of the variables
$y^1$ and $y^2$, which have a much simpler commutation relation than
the original coordinates $x^1$ and $x^2$.

The inverse transformation is easy to find,
\begin{equation}\label{eq:inverse1}
\left\{
\begin{array}{ccl}
x^1 &=&  y^1 \,+\, \varphi \big( t(y^1) \; y^2 \big)\\
x^2 &=& t(y^1) \; y^2  \;,
\end{array}
\right.
\end{equation}
where all the products are commutative. Therefore, we also have
the direct transformation:
\begin{equation}\label{eq:direct}
\left\{
\begin{array}{ccl}
y^1 &=&  x^1 \,-\, \varphi(x^2)\\
y^2 &=& x^2 / t(x^1 \,-\, \varphi(x^2))\;.
\end{array}
\right.
\end{equation}

An  important tool in the construction of noncommutative field
theories is the definition of derivatives. Since we know two
variables $y^1$ and $y^2$, which verify a `canonical' (i.e.,
constant-$\theta$) commutation relation, it is natural to
construct derivative operators in terms of them:
\begin{eqnarray}\label{eq:defder}
D_1 &\equiv& \frac{\partial}{\partial y^1} \,=\,
\frac{\partial}{\partial x^1} \, +\, x^2 \, \frac{\partial t(x^1 -
\varphi(x^2))/ \partial x^1}{t(x^1 - \varphi(x^2))} \, \big[
\varphi'(x^2) \frac{\partial}{\partial x^1} +
\frac{\partial}{\partial x^2}
\big]\nonumber\\
D_2 &\equiv& \frac{\partial}{\partial y^2}\,=\,  t(x^1 -
\varphi(x^2)) \, \big[ \varphi'(x^2) \frac{\partial}{\partial x^1}
+ \frac{\partial}{\partial x^2} \big] \;.
\end{eqnarray}

We can then obtain an explicit expression for the $\star$-product
between $f$ and $g$: from (\ref{cinco}) and (\ref{eq:direct})
using the derivatives above defined,
\begin{equation}\label{eq:newstar}
(f \star g) (x^1,x^2) \;=\; {\rm{exp}}\frac{i}{2} \,
\big(\theta(\tilde x ^1,\tilde x ^2){D_1}{\tilde D}_2-\theta(x
^1,x ^2)D_2 {\tilde D}_1
 \big)\,f(x) g(\tilde x)
\; \Big \vert _{\tilde x = x} \;,
\end{equation}
where:
\begin{equation}
\theta(x^1,x^2)\;=\;   \theta_0 \,+\, \delta \, \tanh \big(
\frac{x^1 - \varphi(x^2)}{\epsilon} \big) \;.
\end{equation}
This expression shows explicitly how the interface enters in the
star product.

It may be hard to attempt a direct proof of the fact that the
operators defined in this way do satisfy Leibniz rule (with the
star product (\ref{eq:newstar})).  However, this can be easily
achieved if we note that $D_i$ are in fact {\em inner\/}
derivations. Indeed,
\begin{equation}
 \frac{\partial}{\partial y^1} f(y)\,=\,i
\theta_0 [y^2,f(y)]_*\,=\,\,i \theta_0 [x^2 \,
t^{-1}(x^1-\varphi(x^2)),f(y(x))]_\star\;,
\end{equation}
and a similar expression for $D_2$. See, for example, \cite{ft}
for a discussion of inner derivations in a general setting.
\subsection{Noncommutative Field Theory}\label{ssec:ncft}

One immediate outcome  of (\ref{eq:direct}) is the expression for
the integration measure, assumed to be defined for $y_1$ and
$y_2$, in terms of $x^1$ and $x^2$:
\begin{equation}\label{eq:measure}
d\mu \;\equiv\; dy^1 \, dy^2 \;=\; dx^1 \, dx^2 \, |\frac{\partial
(y^1,y^2)}{\partial (x^1 , x^2)}| \;\equiv\; dx^1 \, dx^2 \,
|J|\;.
\end{equation}
where
\begin{equation}
J\;=\;\big[ t(x^1 -\varphi(x^2)\big]^{-1} \;.
\end{equation}

It is now evident that the case in which $\theta$ changes its sign
through the interface is qualitatively different: the change of
variables becomes singular on ${\mathcal C}$, when $x^1 \,=\,
\varphi (x^2)$, and as a consequence one has to introduce
different changes of variables on different charts.

We have at this stage the complete set of tools to define a
noncommutative field theory. For a scalar field in Euclidean time
$\sigma(\tau,x^1,x^2)$, satisfying a reality condition which we
will determine in a moment, we may define the action:
\begin{equation}\label{eq:defs}
S[\sigma] \;=\; \int\frac{d\tau dx^1 dx^2}{|t(x^1 - \varphi(x^2))|} \,
\Big[\frac{1}{2}\big( \partial_\tau\sigma \star  \partial_\tau\sigma + D_j \sigma \star D_j \sigma + m^2 \sigma \star \sigma \big)
+ V_\star (\sigma) \Big]
\end{equation}
where the $D_j$'s ($j = 1,2$)  have been defined in (\ref{eq:defder}), and the
$\star$-product is the one of (\ref{eq:newstar}). This expression may of
course be converted into its equivalent version in the variables $y^j$,
a procedure that yields the much simpler expression:
\begin{equation}\label{eq:defsq}
S[{\tilde \sigma}] \;=\; \int d\tau dy^1 dy^2 \, \Big[\frac{1}{2}\big(
\partial_\tau{\tilde \sigma} \ast  \partial_\tau{\tilde \sigma}  + \partial_j{\tilde \sigma} \ast  \partial_j{\tilde \sigma}+ m^2
{\tilde \sigma} \ast   {\tilde \sigma}\big) + V_\ast  ({\tilde \sigma}) \Big]
\end{equation}
with ${\tilde \sigma}(\tau, y) = \sigma(\tau,x(y))$, and
$\partial_j \equiv \partial/{\partial y^j}$. Here it becomes
evident that a positive action requires $\tilde \sigma$ to be a
real function of $(\tau, y)$; then, noting that the transformation
(\ref{eq:direct}) is real, $\sigma(\tau,x)$ has to be real.

Working in terms of the variables $(y^1,y^2)$ provides another
important simplification: from the Moyal product
(\ref{eq:moyal1}), we see that
\begin{equation}
\int dy^1 dy^2 \, f(y) * f(y) \,=\,\int dy^1 dy^2 \, f(y)^2
\;\;\; \forall \; f(y) \;.
\end{equation}
 Then the quadratic part of the action
coincides with its commutative counterpart. Noncommutative effects
appear in $V_\star$ and, when we return to the original
coordinates $(x^1,x^2)$, they show up in the existence of
nontrivial metric and derivatives.

This procedure of `pulling back' the action to the coordinates
$y^j$ may be used, for example, to derive the free propagator in
terms of the `physical' coordinates $x^j$. Considering for
simplicity the massless case, we see that:
\begin{eqnarray}
\langle\sigma(\tau,x) \sigma(\tau',x')\rangle &=& \frac{1}{4 \pi} \Big[ (\tau - \tau')^2 +
(x^1 - x'\,^1 - \varphi(x^2) + \varphi(x\,'^2))^2 \nonumber\\
&+& \big(\frac{x^2}{t(x^1 \,-\, \varphi (x^2))} - \frac{x'\,^2}{
t(x'\,^1 \,-\, \varphi (x'\,^2)) } \big)^2 \Big]^{-1/2}\;.
\end{eqnarray}

This expression allows one to derive, in particular, the
propagator for points on the interface. Indeed, considering the
propagator for the two points $x=(x^1,x^2)$ and
$x'=(x'\,^1,x'\,^2)$, where $x^1 = \varphi(x^2)$ and $x'\,^1 =
\varphi(x'\,^2)$, we see that:
\begin{equation}
\langle\sigma(\tau,x) \sigma(\tau',x')\rangle|_{\mathcal C} \;=\;
\frac{1}{4 \pi} \Big[ (\tau - \tau')^2 + (x^2 - x'\,^2)^2
\Big]^{-1/2}\;,
\end{equation}
where we used the fact that $t=1$ on ${\mathcal C}$.  The result
is different for pairs of arguments that are entirely inside each
one of the two constant-$\theta$ regions. For example, for the
case $x^1 >> \varphi(x^2)$, $x'\,^1 >> \varphi(x'\,^2)$, we have:
\begin{equation}
\langle\sigma(\tau,x) \sigma(\tau',x')\rangle \;\sim\; \frac{1}{4
\pi} \Big[ (\tau - \tau')^2 + (x^1 - x'\,^1)^2 \,+\,
(\frac{\theta_0}{\theta_+})^2 \, (x^2 - x'\,^2)^2 \Big]^{-1/2}\;.
\end{equation}
and a similar expression (with $\theta_+ \leftrightarrow \theta_-$) for the other region. Thus
the presence of the interface introduces a fundamental change in the
propagator, which never becomes equal to the free one, except for
points on the interface.

We have mentioned the difficulties that arise when $\theta$
vanishes; indeed, it is clear now that this implies a singularity
in the change of variables. Assuming that $\theta_-$ and
$\theta_+$ are such that $\theta$ vanishes on the interface, we
know that $t$ will vanish (with a non-zero normal derivative) at
the interface. One can, however, still define an action for the
region where $\theta = \theta_+$, say. It is easy to write that
action in terms of the variables $y^j$, and then pass to the
physical variables whenever necessary:
\begin{equation}\label{eq:defsq+}
S[{\tilde \sigma}] \;=\; \int_{y_1 >0} d\tau dy^1 dy^2 \, \Big[\frac{1}{2}
\big(
\partial_\tau{\tilde \sigma}* \partial_\tau{\tilde \sigma}  +
\partial_j{\tilde \sigma} * \partial_j{\tilde \sigma}+ m^2
{\tilde \sigma} *  {\tilde \sigma}\big) + V_* ({\tilde \sigma})
\Big] \;,
\end{equation}
since $y^1 > 0$ amounts to $x^1 > \varphi(x^2)$.

It is obvious that the existence of a border at $y^1 = 0$ requires the
introduction of a boundary condition for ${\tilde \sigma }$. A non-trivial
consequence of the non-linear relation between the $y$'s and the physical
variables is that the Neumann condition: $\partial_1{\tilde \sigma}(0,y_2) = 0$
becomes too strong in terms of the new variables:
\begin{equation}
 \Big[  \frac{\partial \sigma }{\partial x^1} \, +\, x^2
\, \frac{\partial t(x^1 - \varphi(x^2))/\partial x^1}{t(x^1 - \varphi(x^2))} \,
\big( \varphi'(x^2) \frac{\partial \sigma }{\partial x^1} + \frac{\partial \sigma }{\partial x^2}\big) \Big]|_{x^1
\to \varphi(x^2)} \;=\; 0\;.
\end{equation}
Thus we should have not only $\frac{\partial \sigma }{\partial x^1}=0$, but also
\begin{equation}
\varphi'(x^2) \frac{\partial \sigma }{\partial x^1} + \frac{\partial \sigma }{\partial x^2} =0
\end{equation}
at the interface, since $t$ vanishes with a non-zero derivative
there. The last condition means that the derivative of $\sigma$in
the direction of the curve ${\mathcal C}$ should vanish, since
$(\varphi'(x^2), 1)$ is tangent to that curve.

\section{Generalization of the method}\label{sec:general}
We now turn to generalizations of the change of variables approach
to deal with the general situation~\footnote{In this section we
will not continue to interpret $\theta(x)$ as arising from a
nontrivial metric. Moreover, we shall  not distinguish between sub
and super-indices.}
\begin{equation}
  \label{eq:gencommrel}
[ x_1 , x_2 ]_\star \,=\,i \, \theta(x_1 , x_2)\;.
\end{equation}
One may wonder to what extent the method and results of sections
\ref{sec:metric} and \ref{sec:interface} depend on choosing the
Moyal product to represent relation (\ref{cuatro}), which is
equivalent to the Weyl ordering prescription to define the map
between operators and functions. To show that the change of
variables is not restricted to this case, we will now work in a
normal order, which arises naturally from the holomorphic
representation of two-dimensional systems.  This framework is
particularly appropriate to construct  scalar multisolitons,   as
described in \cite{GMS} and extended  in \cite{SV}, where the
generalization of Berezin approach to deformation quantization is
applied to the case of arbitrary Kh\"aler manifolds.

\subsection{The holomorphic representation}\label{holo}

Planar systems are characterized by creation and annihilation
operators ${\hat a}$ and $\hat a ^\dag$ that satisfy
\begin{equation}
[\hat a , \hat a ^\dag]\,=\,1 \,,
\end{equation}
and we shall use that representation to discuss different changes
of variables. For future convenience, we symbolize them as
\begin{equation}
\hat a \to \hat{\bar z} \;\;\;,\;\;\; \hat a ^\dag\to \hat z \;.
\end{equation}

We represent the algebra over the Hilbert space $\mathcal H$ with
basis $\{\vert z \rangle \}$ of eigenstates of $\hat z$ (coherent
states). On this basis, the action of the operators is
\begin{equation}
\hat z  \to z \;\;\;,\;\;\; \hat {\bar z} \to
\frac{\partial}{\partial z} \;. \end{equation} The scalar product
in $\mathcal H$ is defined by
\begin{equation}
  \label{eq:scalarprod}
\langle f  \vert g \rangle \equiv \int \frac{dz \, d \bar z}{2\pi
i}\; {\rm e} ^{-\bar z z} \; \bar f (\bar z)\, g(z) \;,
\end{equation}
with
\begin{equation}
g(z)\equiv \langle z  \vert g \rangle \;\;\;,\;\;\;  \bar f (\bar
z) \equiv \langle f  \vert \bar z \rangle \;.
\end{equation}
With this scalar product, $z$ and $\partial / \partial z$ are
relatively adjoint, and the eigenstates
\begin{equation}
\langle z  \vert n \rangle \,=\,\frac{z^n}{\sqrt{n !}}
\end{equation}
of the Hermitean operator
\begin{equation}
z \partial_z \,=\, a ^ \dag\, a \end{equation} form an orthonormal
basis. Besides, we can obtain the integral representation of the
identity operator:
\begin{equation}
  \label{eq:intrepr}
f(z)\,=\,\sum_n \langle  z  \vert n \rangle\,\langle n  \vert f
\rangle\,=\,\int \frac{dz \, d \bar z}{2\pi i}\; {\rm e} ^{-\bar z
' z'}\, {\rm e} ^{-\bar z ' z} \, f(z')\;.
\end{equation}

Now we represent the algebraic structure of $\mathcal A$ in the
space of functions $f(z, \bar z)$ by defining a $\star$-product.
To each operator $A(\hat z , \hat{\bar z})$ we associate the
function
\begin{equation}
  \label{eq:Smap}
A(z, \bar z)\equiv \mathcal S [A(\hat z , \hat{\bar z})] \equiv
\frac{\langle z \vert A(\hat z , \hat{\bar z})\vert \bar z
\rangle}{\langle z \vert \bar z \rangle} \;.
\end{equation}
This corresponds to the normal-ordering prescription in which all
the $\hat z$ are to the left of all the $\hat{ \bar z}$. We define
the $\star$-product as
\begin{equation}
  \label{eq:starz}
(A \star B) (z, \bar z)\,=\, \mathcal{S} [\mathcal{S}^{-1}(\hat A)
\mathcal{S}^{-1} (\hat B)] \;.
\end{equation}
Then
\begin{equation}
  \label{eq:starz2}
(A \star B) (z, \bar z)\,=\,\int \frac{dz' \, d \bar z '}{2\pi
i}\; {\rm e} ^{-(z-z')(\bar z - \bar z ')} \; A(z, \bar z ') \,
B(z, \bar z ') \;.
\end{equation}

It is immediate to verify that $\star$-product reproduces the
algebraic structure of $\mathcal A$:
\begin{equation}
  \label{eq:fundrelation}
[\bar z , z]_\star\,=\,1 \;.
\end{equation}
Some useful relations are:
\begin{equation}
  \label{eq:rel1}
A(z) \star B(z, \bar z)\,=\,A(z) B(z, \bar z) \;,
\end{equation}
\begin{equation}
  \label{eq:rel2}
A(z, \bar z) \star B(\bar z)\,=\,A(z,\bar z) B(\bar z) \;,
\end{equation}
\begin{equation}
  \label{eq:rel3}
A(\bar z) \star B(z)\,=\,A(\bar z + \partial_z) B(z) \;,
\end{equation}
where $\partial_z \equiv \partial / \partial z$ is the usual
derivative.

The integral is defined as
\begin{equation}
  \label{eq:integrz}
\int d \mu (z, \bar z ) \, A(z,\bar z) \equiv {\rm Tr} A(\hat z,
\hat {\bar z})\,=\,\int \frac{dz \, d \bar z}{2\pi i}\;A(z, \bar
z)\;.
\end{equation}
The usual derivatives $\partial_z$, $\partial_{\bar z}$ are
realized as $\star$-product commutators,
\begin{equation}
  \label{eq:derivz}
\partial_z \,=\,[\bar z ,\;]_\star \;\;\; ,  \;\;\; \partial_ {\bar z} \,=\, - [z, \;]_\star \;.
\end{equation}

\subsection{General change of variables}\label{general}

Returning to Eq. (\ref{eq:gencommrel}), it is useful to move to
the complex plane $(w, \hat w)$:
\begin{equation}
  \label{eq:complexw}
w\,=\,\frac{1}{\sqrt 2}\,(x_1+i x_2) \;\;\;, \;\;\;\bar
w\,=\,\frac{1}{\sqrt 2}\,(x_1-i x_2) \;.
\end{equation}
Thus we have to study the commutation relation
\begin{equation}
  \label{eq:commrelw}
[w, \bar w]_\star\,=\,\theta\big(\frac{1}{\sqrt 2}\,(w+\bar
w),-\frac{i}{\sqrt 2}\,(w-\bar w)\big) \;.
\end{equation}

We want to define new coordinates $(z,\bar z)$:
\begin{equation}
  \label{eq:change1}
w\,=\,g(z,\bar z)\;\;\; , \;\;\;\bar w\,=\,\bar g (\bar z,z)\; \;,
\end{equation}
such that $[\bar z , z]_\star\,=\,1 $; i.e., the unknown function
$g$ must satisfy
\begin{equation}
  \label{eq:condition1}
g(z, \bar z) \star \bar g (\bar z,z)-\bar g (\bar z,z) \star g(z,
\bar z) \,=\, \tilde \theta (z, \bar z) \;,
\end{equation}
where $\tilde \theta$ is the function $\theta$ expressed in terms
of  $(z,\bar z)$.

If this is the case, then we can still apply the formulae from the
previous section. For instance
\begin{equation}
  \label{eq:star3}
(A \star B) (x_1, x_2)\,=\, (\tilde A \star \tilde B) (z, \bar
z)\,=\,\int \frac{dz' \, d \bar z '}{2\pi i}\; {\rm e}
^{-(z-z')(\bar z - \bar z ')} \; \tilde A(z, \bar z ') \, \tilde
B(z', \bar z) \;,
\end{equation}
where $$\tilde A(z, \bar z ) \equiv A \big(x_j (z , \bar z) \big)
\;$$ The integrals will be of the form
\begin{equation}
  \label{eq:integr2}
I\,=\,\int d \mu (x) \, A(x_1,x_2)
\end{equation}
and comparing with Eq.(\ref{eq:integrz}), we obtain the
integration measure
\begin{equation}
  \label{eq:measure2}
d\mu (x) \,=\, \frac{1}{ 2 \pi i} J(x_1,x_2) \,dx_1 \,dx_2 \,
\end{equation}
where $J$ denotes the Jacobian
\begin{equation}
J(x_1,x_2) \equiv \Big \vert \frac{\partial(z)}{\partial (w)} \Big
\vert (x_1,x_2) \;.
 \end{equation}

Finally, from (\ref{eq:derivz}), two possible inner derivations
are
\begin{equation}
  \label{eq:deriv1}
D_1 A(x)\equiv [\bar f (\bar w , w)\,,\,A(x)]_\star
\,=\,\partial_z \tilde A(z, \bar z) \;,
\end{equation}
\begin{equation}
  \label{eq:deriv2}
D_2 A(x)\equiv- [f (w, \bar w)\,,\,A(x)]_\star \,=\,\partial_{\bar
z} \tilde A(z, \bar z) \;,
\end{equation}
being $f(w,\bar w)$ the inverse function of $g(z, \bar z)$. They
satisfy Leibniz rule and
\begin{equation}
\int d \mu (x) D_j A(x) \,=\,0 \;.
 \end{equation}

\subsection{Conformal Transformations}\label{conformal}

Let us consider the special and important case of conformal
transformations
\begin{equation}
  \label{eq:conftran}
w\,=\,g(z)\;\;\;,\;\;\; \bar w \,=\, \bar g (\bar z) \;,
\end{equation}
with $g$ an analytic function. Applying
Eqs.(\ref{eq:rel1})-(\ref{eq:rel3}) to our present situation, we
see that $g$ must satisfy the condition
\begin{equation}
  \label{eq:condition2}
g(z) \bar g (\bar z)-\bar g (\bar z+\partial_z) g(z) \,=\, \tilde
\theta (z, \bar z) \;.
\end{equation}
This is of course much simpler than the general condition
(\ref{eq:condition1}), but it will only be useful to deal with
some special cases.

For instance, we use this to study the rotation-invariant case
\begin{equation}
  \label{eq:commrelrot}
[ x_1 , x_2 ]_\star \,=\,i \frac{\theta_0}{2} (x_1^2 + x_2 ^2)\;.
\end{equation}
In the complex plane,
\begin{equation}
  \label{eq:wcommrot}
[w, \bar w]_\star\,=\,\theta_0 \, w \bar w \;.
\end{equation}
From the condition  (\ref{eq:condition2}), $w=g(z)$ has to satisfy
\begin{equation}
  \label{eq:conditionrot}
\bar g (\bar z+\partial_z) g(z) \,=\, (1-\theta_0) \, \bar g (\bar
z) g(z) \;.
\end{equation}
This suggests to take
\begin{equation}
  \label{eq:changerot}
w\,=\,g(z)\,=\, l \, {\rm e}^{a z}
\end{equation}
where $l$ is a parameter with units of length and $a$ is a
dimensionless constant. Indeed,
\begin{equation}
\bar g (\bar z+\partial_z) g(z)\,=\, {\rm e}^{a^2 } \,\bar g (\bar
z) g(z) \;;
 \end{equation}
so, $a$ must satisfy
\begin{equation}
 {\rm e}^{a^2 }\,=\,1-\theta_0 \;.
 \end{equation}

According to (\ref{eq:measure2}), the integrals will be of the
form
\begin{equation}
  \label{eq:integralrot}
I\,=\,\int dx_1 \, dx_2 \, (x_1^2 + x_2 ^2)^{-1} \; A(x_1,x_2) \;.
\end{equation}
Since
\begin{equation}
z\,=\, {\rm ln}^{-1/2}(1-\theta _0) \,  {\rm ln}\frac{w}{l} \,
 \end{equation}
the derivations are, using the chain rule,
\begin{equation}
  \label{eq:deriv1rot}
D_1 A(x)\,=\,\partial_z \tilde A (z, \bar z)\,=\,{\rm
ln}^{1/2}(1-\theta _0) \, (x_1+ix_2)\,\Big(
\frac{\partial}{\partial x _1}-i\frac{\partial}{\partial x _2}
\Big)A(x)
\end{equation}

\begin{equation}
  \label{eq:deriv2rot}
D_2 A(x)\,=\,\partial_{\bar z} \tilde A (z, \bar z)\,=\,{\rm
ln}^{1/2}(1-\theta _0) \, (x_1-ix_2)\,\Big(
\frac{\partial}{\partial x _1}+i\frac{\partial}{\partial x _2}
\Big)A(x)
\end{equation}

From
\begin{equation}
x_1 \,=\, \frac{l}{\sqrt 2} \, ( {\rm e}^{a z}+ {\rm e}^{a \bar
z}) \;\;\;\;, \;\;\; x_2 \,=\, -\frac{l}{\sqrt 2} \, i ( {\rm
e}^{a z}- {\rm e}^{a \bar z})
 \end{equation}
and Eqs.(\ref{eq:rel1})-(\ref{eq:rel3}), the fundamental
$\star$-products are
\begin{equation}
x_j \star x _j \,=\, x_j ^2 - \frac{\theta _0}{4} (x_1^2 + x_2 ^2)
 \end{equation}
(no summation over repeated indices) and
\begin{equation}
x_j \star x _k \,=\,x_j x_k+i\frac{\theta _0}{4} \epsilon_{j k}
(x_1^2 + x_2 ^2)
 \end{equation}
for $j \neq k$. It is worth noting that with this star product,
\begin{equation}
x_1 \star x _1 + x_2 \star x _2 \,=\, \big(1-\frac{1}{2} \theta_0
\big)(x_1^2 + x_2 ^2)
 \end{equation}
is also rotation-invariant.

\section{Summary and discussion}\label{sec:conc}
 We have shown how, in some special cases, planar noncommutative
theories with a space-dependent $\theta$ can be equivalently
described in terms of a constant $\theta$, by performing a
suitable change of variables. In this way, we have been able to
find an explicit representation of the noncommutative product, and
to deal with its associated calculus (i.e., derivatives and
integration).

We applied that procedure in section~\ref{sec:metric} to the
concrete example of a $\theta(x)$ depending on only one
coordinate. We have also unravelled  the connection between
space-dependent noncommutative parameter $\theta(x)$  and
associative noncommutativity in a non-trivial background metric
$g_{ij}$. As explained at the end of this section, apart from its
interest {\it per se}, this connection can be useful in the
solution of selfduality  noncommutative soliton equations which
can be derived, one from the other if one introduces an
appropriate background metric.

The  resulting $\star$-product  was later on
(section~\ref{sec:interface}) adapted to the example of a smooth
interface $\mathcal C$ dividing two regions with different
constant $\theta$-values. The resulting noncommutative field
theory obtained in this way could be used, for example, to study
the physics of a reduced (relativistic) lowest Landau level
description, when the magnetic field is not homogeneous, but
rather has two different constant values in two respective regions
(which have $\mathcal C$ as a boundary).

Finally, in section~\ref{sec:general}, we sketched the extension
of the change of variables to more general situations, by taking
advantage of the holomorphic representation for planar systems.
The simple and yet important case in which the redefinition of
variables is realized  by a conformal transformation, has been
analyzed in~\ref{conformal}, applying the results to the
particular case of a rotation-invariant $\theta$.

A crucial point in the method is, of course, to find an adequate
change of variables. This, in turn, depends on the explicit form
of $\theta(x)$, in particular on its symmetries. That is the
reason why it does not seem  possible to construct an {\em
explicit\/} coordinate transformation to address the general case.
Nevertheless, we believe that the present results can contribute
to an understanding of the properties of theories with a
space-dependent noncommutativity in a simpler way, by mapping them
to systems equipped with the standard Moyal product.

\section*{Acknowledgements}
D.~H.~C.~, C.~D.~F and G.~T.~ have been  supported by a
Fundaci\'on Antorchas grant. D.~C and F.~A.~S.~ have been
partially supported by UNLP, CICBA, CONICET and ANPCYT (PICT grant
03-05179).

\end{document}